\documentclass[a4paper,twocolumn,11pt,accepted=2025-05-18]{quantumarticle}
\pdfoutput=1
\usepackage[utf8]{inputenc}
\usepackage[english]{babel}
\usepackage[T1]{fontenc}
\usepackage{graphicx}
\usepackage{dcolumn}
\usepackage{bm}
\usepackage{physics}
\usepackage[x11names]{xcolor}
\usepackage{tikz}
\usepackage{pgfplots}
\pgfplotsset{compat=1.18}
\usetikzlibrary{decorations.pathmorphing}
\usepackage[numbers,sort&compress]{natbib}
\usepackage{hyperref}

\begin{document}


\title{Witnessing mass-energy equivalence with trapped atom interferometers}

\author{Jerzy Paczos}
\email{jerzy.paczos@fysik.su.se}
\affiliation{Department of Physics, Stockholm University, SE-106 91 Stockholm, Sweden}

\author{Joshua Foo}
\email{jfoo@stevens.edu}
\affiliation{Centre for Quantum Computation and Communication Technology, School of Mathematics and Physics, University of Queensland, St. Lucia, Queensland 4072, Australia}
\affiliation{Department of Physics, Stevens Institute of Technology, Castle Point Terrace, Hoboken, New Jersey 07030, USA}
\affiliation{Department of Physics and Astronomy, University of Waterloo, Waterloo, Ontario, Canada, N2L 3G1}

\author{Magdalena Zych}
\email{magdalena.zych@fysik.su.se}
\affiliation{Department of Physics, Stockholm University, SE-106 91 Stockholm, Sweden}
\affiliation{Centre for Engineered Quantum Systems, School of Mathematics and Physics, The University of Queensland, St. Lucia, Queensland, 4072, Australia}

\begin{abstract}
We propose an experimental setup to probe the interplay between the quantum superposition principle and gravitational time dilation arising from the mass-energy equivalence. It capitalizes on state-of-the-art atom interferometers that can keep atoms trapped in a superposition of heights in Earth's gravitational field for exceedingly long times, reaching the minute scale. Our proposal consists of adding two additional laser pulses to the existing experiments that would set up a clock trapped at a superposition of heights, reading a quantum superposition of relativistic proper times. We develop a method to include relativistic corrections to Bloch oscillations, which describe the trapped part of the interferometer. We derive the trajectories and corresponding phases acquired in each arm of the interferometer. We then show that a superposition of proper times manifests in the interference pattern in two ways: visibility modulations and a shift of the atom’s resonant frequency. We argue that the latter might be observable with current technology.

\end{abstract}

\maketitle


%
%
\noindent {\it ---Introduction.} A key prediction of general relativity is that time is not a global background parameter but flows at different rates depending on the spacetime geometry, a phenomenon known as time dilation~\cite{einstein1905}. The physical effects produced by time dilation have been tested in numerous paradigmatic experiments: Pound and Rebka were the first to observe an altitude-dependent gravitational redshift in gamma rays emitted between the top and bottom of a tower~\cite{Pound1959}; Hafele and Keating directly tested special and general relativistic time dilation by comparing atomic clocks at different heights and moving at different speeds~\cite{Hafele1972}; and Shapiro proposed to measure the reduction in the speed of light for electromagnetic waves traveling across regions subject to a gravitational potential~\cite{Shapiro1964}, an effect later observed in~\cite{Shapiro1971}. These effects are fully explainable in classical relativistic physics.

The first experiment to measure the effect of gravity on the quantum wavefunction of a single particle was performed by Colella, Overhauser, and Werner (COW)~\cite{Colella1975}. In this setup, neutrons travel in a superposition of heights in Earth’s gravitational field, following the geometry of a Mach-Zehnder-type interferometer. The neutrons, traversing arms at different heights in the gravitational potential, acquire a relative phase. The resulting interference pattern was measured, yielding a phase shift consistent with Earth’s gravitational potential. These experiments, and related measurements with atoms~\cite{Kasevich1991, Kasevich1992, Peters1999, McGuirk2002, Fixler2007, Lamporesi2008, Rosi2014, Rosi2015, Kovachy2015, Hu2017, Asenbaum2020, Charriere2012, Andia2013, Zhang2016, Xu2019, Panda2024}, thus test non-relativistic gravitational effects in quantum systems.

The proposal to unambiguously probe the interplay between quantum mechanics and general relativity was developed in~\cite{Sinha2011, Zych2011}. These works considered interferometric experiments with particles having an internal degree of freedom that can be treated operationally as a ``quantum clock''. The internal evolution would then depend on the proper time along the paths taken through curved spacetime~\cite{Sinha2011, Zych2011, Zych2012, Castro-Ruiz2017, Loriani2019, Roura2020, Castro-Ruiz2020, Smith2020, Khandelwal2020, Grochowski2021, Paczos2024, Debski2024}. Specifically, the authors of~\cite{Zych2011} predicted that the experiment would exhibit oscillations in the interference visibility --- defined as the contrast between constructive and destructive fringes --- which, in non-relativistic COW-type treatments, is always maximal in principle. The predicted periodic modulations of the visibility would be an effect that arises unambiguously from the interplay between relativistic and quantum physics. Several proposals have been made to investigate this effect in interferometric experiments with electrons~\cite{Bushev2016}, and with atoms~\cite{Loriani2019, Roura2020, Roura2021, Ufrecht2020, Meltzer2024}. However, given the size of the interferometer required for a reasonable visibility drop~\cite{Zych2011}, it is rather unlikely that this effect will be observed in the near future.

Derivations of the dynamics of relativistic quantum particles with quantized internal mass-energy --- necessary for predicting the above experiments --- have been performed both within relativistic quantum mechanics~\cite{Sonnleitner2018, Zych2019, Schwartz2019_1, Schwartz2019_2} and via quantum field theory on curved backgrounds~\cite{Zych2017, Zych2018, Roura2020}. In the low-energy limit for the center-of-mass motion of the composite particles, these two approaches coincide. These results have been applied to model particle dynamics in clock interferometers~\cite{Bushev2016, Loriani2019, Roura2020, Roura2021, Pumpo2021, Ufrecht2020} and in trapped-particle setups~\cite{KrauseLee2017, Tobar2022}. In particular, they have been used in optical-clock studies to explore relativistic contributions to the clock's frequency and precision in ~\cite{Yudin2018, Haustein2019, Lahuerta2022}, and, most recently, to explore associated many-body effects~\cite{Chu2024}.

In this work, we propose a concrete experimental setup for witnessing time dilation effects in atom interferometry. Our idea is inspired by the rapid development of trapped atom interferometers~\cite{Charriere2012, Andia2013, Zhang2016, Xu2019, Panda2024} utilizing Bloch oscillations to keep atoms in a coherent superposition of heights in a gravitational field for increasingly long times, reaching the minute scale~\cite{Panda2024}. We propose a modification of the existing interferometers by adding two ``clock pulses'' at the beginning and end of the Bloch oscillations, effectively placing a clock in a superposition of heights and thus of proper times. We develop a method to include relativistic mass–energy corrections in the Bloch oscillations and use it to derive all phases acquired along the interferometer trajectories. We then analyze the resulting interference pattern. We recover the visibility modulations predicted in~\cite{Zych2011}, which assumed fixed classical trajectories of the clocks, and crucially predict a clock frequency shift that renders this proposal experimentally feasible.

Indeed, the proposed setup implements Ramsey spectroscopy for spatially superposed atoms, allowing for measurements of their resonant frequencies. We show that relativistic effects manifest therein as resonant-frequency shifts. We argue that this will be easier to observe than the visibility oscillations thanks to the extreme precision of frequency measurements~\cite{Bothwell2022, Zheng2022}. For the height separations reported in ref.~\cite{Zhang2016}, the predicted fractional frequency shift is of order $10^{-22}$ --- one order of magnitude smaller than current experimental precision. By revealing effects potentially detectable with existing technology, our proposal represents a major step toward experimental tests of the interplay between quantum mechanics and general relativity.

%
%

\begin{figure}
    \centering
    \begin{tikzpicture}[scale=0.6]
        \draw[red, thick, domain = -7.1:-3.5, variable = \x]  plot ({\x},{-0.25*(\x+3.5)*(\x+3.5)-0.35});
        \draw[blue, thick, domain = -4.9:-2, variable = \x]  plot ({\x},{-0.25*(\x+2)*(\x+2)+1.25});
        \draw[blue, thick, domain = -3.5:-2, variable = \x]  plot ({\x},{-0.25*(\x+2)*(\x+2)+0.25});
        \draw[red!20, thick, dashed, domain = -3.5:-2.5, variable = \x]  plot ({\x},{-0.25*(\x+3.5)*(\x+3.5)-0.35});
        \draw[red!20, thick, dashed, domain = -3.5:-2.5, variable = \x]  plot ({\x},{-0.25*(\x+3.5)*(\x+3.5)+0.7});
        \draw[blue!20, thick, dashed, domain = 3.5:4.25, variable = \x]  plot ({\x},{-0.25*(\x-2)*(\x-2)+1.25});
        \draw[red!20, thick, dashed, domain = 3.5:4.2, variable = \x]  plot ({\x},{-0.25*(\x-0.6)*(\x-0.6)+1.8});
        \draw[red, thick, domain = 3.5:5.7, variable = \x]  plot ({\x},{-0.25*(\x-0.6)*(\x-0.6)+2.8});
        \draw[blue, thick, domain = 2:3.5, variable = \x]  plot ({\x},{-0.25*(\x-2)*(\x-2)+1.25});
        \draw[blue, thick, domain = 2:6, variable = \x]  plot ({\x},{-0.25*(\x-2)*(\x-2)+0.25});
        \foreach \y in {-3,...,3}
        {
            \shade[top color=Green3!60, bottom color=Green3!60, middle color=white] (-2,0.5*\y) rectangle (2,{0.5*(\y+1)});
        }
        \draw[blue, thick] (-2,1.25) -- (2,1.25);
        \draw[blue, thick] (-2,.25) -- (2,.25);
        \draw[domain=-2.75:2.25, smooth, samples=100, Red3] plot ({-2+0.05*sin(20*\x r)}, {\x});
        \draw[domain=-2.75:2.25, smooth, samples=100, Red3] plot ({2+0.05*sin(20*\x r)}, {\x});
        \draw[domain=-2.75:2.25, smooth, samples=100, RoyalBlue3] plot ({-4.9+0.05*sin(20*\x r)}, {\x});
        \draw[domain=-2.75:2.25, smooth, samples=100, RoyalBlue3] plot ({-3.5+0.05*sin(20*\x r)}, {\x});
        \draw[domain=-2.75:2.25, smooth, samples=100, RoyalBlue3] plot ({3.5+0.05*sin(20*\x r)}, {\x});
        \draw[domain=-2.75:2.25, smooth, samples=100, RoyalBlue3] plot ({4.9+0.05*sin(20*\x r)}, {\x});
        \draw (-5.1,2.25) rectangle (-4.7,2.35);
        \draw (-5.1,2.35) rectangle (-4.7,2.5);
        \draw (-5.1,-2.75) rectangle (-4.7,-2.85);
        \draw (-5.1,-2.85) rectangle (-4.7,-3);
        \draw (-3.7,2.25) rectangle (-3.3,2.35);
        \draw (-3.7,2.35) rectangle (-3.3,2.5);
        \draw (-3.7,-2.75) rectangle (-3.3,-2.85);
        \draw (-3.7,-2.85) rectangle (-3.3,-3);
        \draw (-2.2,2.25) rectangle (-1.8,2.35);
        \draw (-2.2,2.35) rectangle (-1.8,2.5);
        \draw (-2.2,-2.75) rectangle (-1.8,-2.85);
        \draw (-2.2,-2.85) rectangle (-1.8,-3);
        \draw (5.1,2.25) rectangle (4.7,2.35);
        \draw (5.1,2.35) rectangle (4.7,2.5);
        \draw (5.1,-2.75) rectangle (4.7,-2.85);
        \draw (5.1,-2.85) rectangle (4.7,-3);
        \draw (3.7,2.25) rectangle (3.3,2.35);
        \draw (3.7,2.35) rectangle (3.3,2.5);
        \draw (3.7,-2.75) rectangle (3.3,-2.85);
        \draw (3.7,-2.85) rectangle (3.3,-3);
        \draw (2.2,2.25) rectangle (1.8,2.35);
        \draw (2.2,2.35) rectangle (1.8,2.5);
        \draw (2.2,-2.75) rectangle (1.8,-2.85);
        \draw (2.2,-2.85) rectangle (1.8,-3);
        \draw [stealth-stealth] (-2,-3.2) -- (2,-3.2);
        \node at (0,-3.6) {$T_{\rm B}$};
        \draw [stealth-stealth] (-4.9,-3.2) -- (-3.5,-3.2);
        \node at (-4.2,-3.6) {$T$};
        \draw [stealth-stealth] (-3.5,-3.2) -- (-2,-3.2);
        \node at (-2.75,-3.6) {$T'$};
        \draw [stealth-stealth] (4.9,-3.2) -- (3.5,-3.2);
        \node at (4.2,-3.6) {$T$};
        \draw [stealth-stealth] (3.5,-3.2) -- (2,-3.2);
        \node at (2.75,-3.6) {$T'$};
    \end{tikzpicture}
    \caption{Proposed setup. All motion is restricted to one dimension (parallel to the gravitational field). Blue wavy lines represent $\pi/2$ Bragg pulses splitting the trajectory of the atom, while red wavy lines represent the $\pi/2$ clock pulses affecting only the internal state of the atom. The optical lattice (green striped pattern) is turned on exactly when the blue trajectories reach their apex and off after time $T_{\rm B}$. Shaded dashed lines represent trajectories lost within the interferometer.}
    \label{Fig::Experimental setup}
\end{figure}
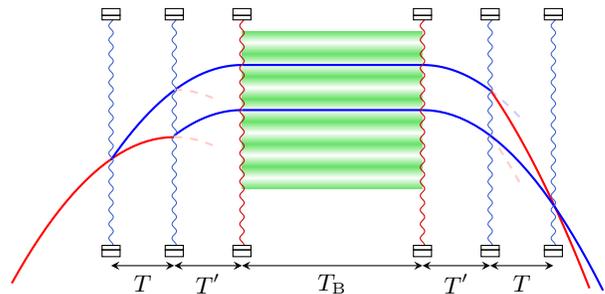

\noindent {\it ---Setup.} Let us consider a one-dimensional atom interferometry setup,  Fig.~\ref{Fig::Experimental setup}, closely resembling configurations implemented in~\cite{Charriere2012, Andia2013, Zhang2016, Xu2019, Panda2024}. It involves highly cooled atoms\footnote{In experiments~\cite{Zhang2016, Xu2019, Panda2024}, the temperature reaches a few hundred nanokelvins.} moving along the gravitational field $g$, with their velocity manipulated by Bragg pulses and an optical lattice. We assume that atoms have a two-level internal structure, with ground and excited states separated by energy~$\hbar\omega_0$.

Atoms are prepared in their ground state and launched upward with initial velocity~$v_0$, entering free fall under gravity. Then, two $\pi/2$ Bragg pulses are applied, separated by an interval~$T$. Each pulse splits the atomic trajectory into two: one retains its velocity, while the other is shifted by~$\Delta v$. Consequently, each initial trajectory splits into four paths with two distinct velocities: two with $v_0-gT$ and two with $v_0-gT+\Delta v$. We focus on the latter two paths, assuming that the others are lost, as in~\cite{Charriere2012, Andia2013, Zhang2016, Xu2019, Panda2024}.

When the relevant trajectories reach their apex at time $T+T'=(v_0+\Delta v)/g$, the optical lattice is switched on, and atoms perform Bloch oscillations. When the lattice is switched on, we introduce our modification—a clock pulse at frequency~$\omega$ that places atoms into an equal superposition of the two internal states without affecting their momentum~\cite{Dicke1953, Grynberg1983, Bergquist1987, Ido2003, Katori2003, Alden2014, Janson2024}. The optical lattice remains on for a duration~$T_\mathrm{B}$, after which the second clock pulse is applied and the lattice is turned off. The atoms then fall freely for time~$T'$, after which another pair of $\pi/2$ Bragg pulses (separated by an interval~$T$) is applied to recombine the trajectories. At this stage, two additional trajectories are lost, unlike in~\cite{Charriere2012, Andia2013, Zhang2016, Xu2019, Panda2024}, where all output ports are measured, although only two interfere (a simplification introduced here for clarity in the subsequent analysis).

In the end, one measures the probabilities $P_g^{(j)}$ that the atoms are found in the ground state in output trajectory $j\in\{0,1\}$ (with 0 and 1 corresponding to the red and blue trajectories in Fig.~\ref{Fig::Experimental setup}, respectively). Apart from the non-relativistic phase $\Delta\phi$ between the upper and lower ground-state trajectories (measured in typical interferometric experiments, e.g.~\cite{Zhang2016}), there will be additional phases $\delta_{\rm d}$ and $\delta_{\rm u}$ acquired within the optical lattice by the excited state on the lower and upper trajectories and which can be measured with the proposed setup. The probability $P_g^{(j)}$ is given by (see Appendix~\ref{supp:probability calculation} for derivation):
\begin{widetext}
    \begin{equation}\label{partial probability}
        P_g^{(j)}=\frac{1}{16}\bigg[1-\cos\left(\frac{\delta_{\rm u}+\delta_{\rm d}}{2}\right)\cos\left(\frac{\delta_{\rm u}-\delta_{\rm d}}{2}\right)+2(-1)^{j}\cos\left(\Delta\phi+\frac{\delta_{\rm u}-\delta_{\rm d}}{2}\right)\sin\left(\frac{\delta_{\rm d}}{2}\right)\sin\left(\frac{\delta_{\rm u}}{2}\right)\bigg].
    \end{equation}
\end{widetext}
To eliminate the $\Delta\phi$-dependence, we calculate the total probability $P_g=P_g^{(0)}+P_g^{(1)}$ of finding the atom in the ground state
\begin{equation}\label{probability}
    P_g=\frac{1}{8}\left[1-\cos\left(\frac{\delta_{\rm u}-\delta_{\rm d}}{2}\right)\cos\left(\frac{\delta_{\rm u}+\delta_{\rm d}}{2}\right)\right].
\end{equation}
This probability depends only on the evolution confined to the optical‑lattice region, where the atom effectively behaves as a clock in a superposition of heights. The presence of two clock pulses is the sole modification to previous implementations~\cite{Charriere2012, Andia2013, Zhang2016, Xu2019, Panda2024}. Note that without these pulses, the ground-state probability would remain fixed at 1/4 (recall that we lose some of the trajectories). In our setup, they oscillate between 0 and 1/4 as a function of the time $T_{\rm B}$ between the clock pulses.

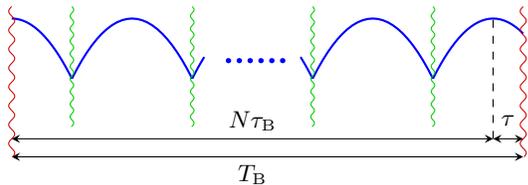
\begin{figure}
    \centering
    \begin{tikzpicture}[scale=0.8]
            \draw[blue, thick, domain = 1:2, variable = \x]  plot ({\x},{-(\x-1)*(\x-1)+2});
            \draw[blue, thick, domain = 2:4, variable = \x]  plot ({\x},{-(\x-3)*(\x-3)+2});
            \draw[blue, thick, domain = 4:4.2, variable = \x]  plot ({\x},{-(\x-5)*(\x-5)+2});
            \draw [blue, line width=2pt, line cap=round, dash pattern=on 0pt off 2\pgflinewidth] (4.6,1.3) -- (5.6,1.3);
            \draw[blue, thick, domain = 5.8:6, variable = \x]  plot ({\x},{-(\x-5)*(\x-5)+2});
            \draw[blue, thick, domain = 6:8, variable = \x]  plot ({\x},{-(\x-7)*(\x-7)+2});
            \draw[blue, thick, domain = 8:9.5, variable = \x]  plot ({\x},{-(\x-9)*(\x-9)+2});
            \draw [stealth-stealth] (1,0) -- (9,0);
            \node at (5,0.3) {$N\tau_{\rm B}$};
            \draw [stealth-stealth] (9,0) -- (9.5,0);
            \node at (9.25,0.3) {$\tau$};
            \draw [stealth-stealth] (1,-0.3) -- (9.5,-0.3);
            \node at (5,-0.6) {$T_{\rm B}$};
            \draw[domain=-0.3:2.2, smooth, samples=100, Red3] plot ({1+0.05*sin(20*\x r)}, {\x});
            \draw[domain=-0.3:2.2, smooth, samples=100, Red3] plot ({9.5+0.05*sin(20*\x r)}, {\x});
            \draw[domain=0.2:2.2, smooth, samples=100, Green3] plot ({2+0.03*sin(30*\x r)}, {\x});
            \draw[domain=0.2:2.2, smooth, samples=100, Green3] plot ({4+0.03*sin(30*\x r)}, {\x});
            \draw[domain=0.2:2.2, smooth, samples=100, Green3] plot ({6+0.03*sin(30*\x r)}, {\x});
            \draw[domain=0.2:2.2, smooth, samples=100, Green3] plot ({8+0.03*sin(30*\x r)}, {\x});
            \draw[dashed] (9,0) -- (9,2);
        \end{tikzpicture}
    \caption{Trajectory followed by the atom (at the lower or upper trajectory) in the optical lattice. We adopt a simplified model of Bloch oscillations in which the atom follows a bouncing trajectory at a fixed height. The total time of Bloch oscillations $T_{\rm B}$ need not be equal to an integer multiple of Bloch periods $N\tau_{\rm B}$. Instead, $T_{\rm B}=N\tau_{\rm B}+\tau$ with $\tau\in(-\tau_{\rm B}/2,\tau_{\rm B}/2)$.}
    \label{Fig::Bloch oscillations}
\end{figure}

%
%
\noindent {\it ---Dynamics and phases.} We now analyze the optical‐lattice stage of the interferometer and the additional phases $\delta_{\rm d}$ and $\delta_{\rm u}$ acquired by the excited state. During this stage, the atom undergoes Bloch oscillations in a superposition of two heights separated by $\Delta z$. We adopt a simplified model of Bloch oscillations (see~\cite{Cadoret2009, Andia2013} and Fig.~\ref{Fig::Bloch oscillations}), in which the atom interacts with the optical lattice only when its downward velocity reaches $v_{\rm B}=\hbar k/m$ (here $k$ is the wave vector of the optical-lattice laser and $m$ is the mass of the ground-state atom). At that point, the atom instantaneously exchanges two photons with the lattice. For an appropriately chosen lattice‐light frequency, this leads to velocity reversal occurring always at the same height, causing the atom to oscillate with period $\tau_{\rm B}=2v_{\rm B}/g$. We will assume that the laser frequency has been chosen exactly this way. Each oscillation thus begins and ends at the apex of a free‐fall trajectory. Over the hold time $T_{\rm B}$, the atom completes $N$ full oscillations, with the final one possibly extended or shortened by an additional time $\tau\in(-\tau_{\rm B}/2,\tau_{\rm B}/2)$, reflecting that $T_{\rm B}$ need not be an integer multiple of $\tau_{\rm B}$.

Crucially, due to the mass difference between the excited and ground states of the atom, the condition for oscillations at the fixed height can be satisfied only for one internal state at a time for a given lattice frequency. If the lattice is tuned to satisfy this condition for the ground state, the excited state will fall below the starting height after each oscillation. However, as shown in Appendix~\ref{supp:trajectory fall}, in viable experimental scenarios, the cumulative fall distance is much smaller than the atom’s thermal de Broglie wavelength. Therefore, the perturbative approach of~\cite{Storey1994} is justified, allowing us to approximate that both internal states follow the same trajectory. This approximation is further validated in Appendix~\ref{supp:non-perturbative calculations}.

The phases $\delta_{\rm d}$ and $\delta_{\rm u}$ comprise two contributions: the laser phase $\omega T_{\rm B}$ acquired by the excited state due to the clock pulses, and a propagation-phase correction arising from the different masses of the ground and excited states. To calculate the latter, we follow~\cite{Storey1994} and write the excited‐state Lagrangian as $L_e=L_g+\delta L$, where $L_g$ is the ground‐state Lagrangian and $\delta L$ is a small correction resulting from the rest‑mass difference between the two states. If we denote this mass difference by $\delta m = \hbar\omega_0/c^2$, the Lagrangian correction is given by:
\begin{equation}
    \delta L=-\delta m\left(c^2+gz-v^2/2\right).
\end{equation}
To calculate the propagation‐phase correction, we integrate $\delta L$ along the atom’s trajectory and divide by $\hbar$. The full derivation is given in Appendix~\ref{supp:phases}; a non‐perturbative calculation is presented in Appendix~\ref{supp:non-perturbative calculations}. This yields the expressions for the phases $\delta_{\rm d}$ and $\delta_{\rm u}$:
\begin{equation}\label{phases}
    \begin{split}
        &\delta_{\rm d}=\left[\omega-\omega_0\left(1-\frac{\langle v^2\rangle}{2c^2}\right)\right]T_{\rm B},\\
        &\delta_{\rm u}=\left[\omega-\omega_0\left(1-\frac{\langle v^2\rangle}{2c^2}+\frac{g\Delta z}{c^2}\right)\right]T_{\rm B},
    \end{split}
\end{equation}
where $\langle v^2\rangle=v_{\rm B}^2/3$ is the mean square speed within a single oscillation.

In principle, mass–energy corrections to the interferometric phases are not the only $\mathcal{O}(1/c^2)$ effects present in our proposed experiment. Following Werner et al.~\cite{Werner2024}, one could also include higher-order corrections from the expansion of the free-fall Lagrangian of the relativistic particle, and $\mathcal{O}(1/c^2)$ corrections to the laser phases arising from elastic interactions (Bragg and Bloch pulses). Because these contributions depend only on the atomic trajectory, and since we approximate that both internal states follow the same path, they modify only the interferometric phase $\Delta\phi$, leaving $\delta_{\rm d}$ and $\delta_{\rm u}$ unmodified. However, $\mathcal{O}(1/c^2)$ corrections to $\Delta\phi$ lie beyond the main focus of this work and will be neglected.

Conversely, any corrections to the clock‐pulse phases would affect $\delta_{\rm d}$ and $\delta_{\rm u}$ if present. Note that the laser‐phase corrections calculated in~\cite{Werner2024} arise from assuming resonant atom-light interactions, with the atom interacting with photons whose laboratory‐frame frequencies depend on its position and velocity. Here, we instead assume the clock pulses operate at a fixed laboratory‐frame frequency, permitting non‐resonant processes. Under this assumption, there are no corrections to the clock‐pulse phases, since the atom interacts with photons at the same frequency regardless of its height or velocity.

\begin{figure}
    \includegraphics[width=0.5\textwidth]{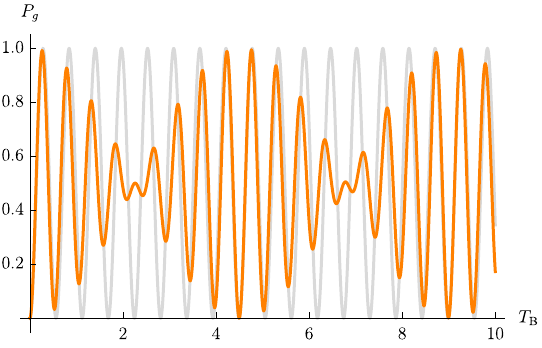}
    \caption{Oscillations of the (normalized) probability $P_g$ for exemplary values $\omega=40\pi$, $\omega_0=1.1\omega$, and $\varepsilon_{\rm k}=0.01$, and two different values of the gravitational correction: $\varepsilon_{\rm g}=0$ (in gray) and $\varepsilon_{\rm g}=\varepsilon_{\rm k}$ (in orange). Note that, apart from visibility modulations, the orange pattern has its frequency shifted against the gray one.}
    \label{fig:interference pattern}
\end{figure}

%
%
\noindent {\it ---Results and discussion.} Let us introduce $\varepsilon_{\rm k}=\langle v^2\rangle/2c^2$ for the kinetic correction in Eq.~\eqref{phases} and $\varepsilon_{\rm g}=g\Delta z/c^2$ for the gravitational correction. Denoting the quantum‑mechanical mean internal frequency by $\langle\omega_0\rangle=\omega_0\left(1-\varepsilon_{\rm k}+\varepsilon_{\rm g}/2\right)$, the ground‑state detection probability (cf.~Eq.~\eqref{probability}) can be written as:
\begin{equation}\label{probability final}
    P_g=\frac{1}{8}\left[1-\mathcal{V}(T_{\rm B})\cos\left(\left[\omega-\langle\omega_0\rangle\right]T_{\rm B}\right)\right],
\end{equation}
where $\mathcal{V}(T_{\rm B}):=\cos\left(\varepsilon_{\rm g}\omega_0T_{\rm B}/2\right)$ is the interferometric visibility. In a generic operating regime away from resonance, the visibility envelope varies slowly with time $T_{\rm B}$ compared to the rapid oscillations of $\cos\left(\left[\omega-\langle\omega_0\rangle\right]T_{\rm B}\right)$. Thus, Eq.~\eqref{probability final} yields an interference pattern comprising fast oscillations modulated by a slowly varying envelope. Apart from the visibility modulations, Eq.~\eqref{probability final} looks like a result of a Ramsey interferometry experiment for an atom with internal transition frequency $\langle\omega_0\rangle$. For a fixed $T_{\rm B}$, scanning the clock‐pulse frequency $\omega$ reveals the resonance at $\omega = \langle\omega_0\rangle$. Consequently, relativistic effects arising from mass–energy equivalence can be probed in two ways: by observing visibility oscillations, and by measuring the resulting fractional frequency shift
\begin{equation}\label{frequency shift}
    \left\langle\frac{\delta\omega_0}{\omega_0}\right\rangle\equiv\frac{\langle\omega_0\rangle-\omega_0}{\omega_0}=-\varepsilon_{\rm k}+\frac{\varepsilon_{\rm g}}{2}.
\end{equation}
Note that whereas visibility oscillations are sensitive only to the gravitational correction $\varepsilon_{\rm g}$, the frequency shift depends on both $\varepsilon_{\rm g}$ and $\varepsilon_{\rm k}$. To distinguish the gravitational and kinetic contributions in fractional‐frequency‐shift measurements, one can modulate the height separation $\Delta z$, which alters $\varepsilon_{\rm g}$ but not $\varepsilon_{\rm k}$. The probability $P_g$ as a function of time is shown in Fig.~\ref{fig:interference pattern} for exemplary values of $\omega$, $\omega_0$, $\varepsilon_{\rm g}$, and $\varepsilon_{\rm k}$.

Assuming the experimental parameters as in~\cite{Zhang2016}, we would have the velocity $v_{\rm B}\sim10\,\text{mm/s}$, and the vertical separation of the trajectories $\Delta z\sim10\,\mu\text{m}$. This would mean that the corrections would be of the order $|\varepsilon_{\rm k}|\sim|\varepsilon_{\rm g}|\sim10^{-22}$. The fractional frequency shift of that order is very close to what can be currently detected~\cite{Hutson2019, Bothwell2022, Zheng2022, Zheng2023}. Indeed, state-of-the-art atomic clocks reach the multiple-measurement precision of the order of $10^{-21}$~\cite{Bothwell2022, Zheng2022, Zheng2023} (the lowest single-measurement precision in such experiments is $\sim10^{-19}$~\cite{Aeppli2024}), which is the order of magnitude for the fractional frequency shift in the proposed experiment for separations $\Delta z\sim100\,\mu\text{m}$, which has already been achieved in~\cite{Zhang2016}, at the expense of shorter coherence time.

On the other hand, for a transition frequency $\omega_0\sim10^{15}\,\mathrm{Hz}$ and a lattice-hold time $T_{\rm B}\sim1\,\mathrm{s}$ (as in~\cite{Zhang2016}), with $|\varepsilon_{\rm g}|\sim10^{-22}$, the predicted visibility drop is of order $10^{-15}$. This is far below current experimental sensitivity. Although strontium atoms can support coherent Bloch oscillations for over $100\,\mathrm{s}$~\cite{Tarallo2012}, the longest achieved to date is $\sim1\,\mathrm{s}$. Minute-scale oscillation times have been demonstrated with cesium atoms~\cite{Xu2019, Panda2024}, but their lower $\omega_0$ renders the effect even weaker. Extending $T_{\rm B}$ to $100\,\mathrm{s}$ would amplify the visibility oscillations by $10^4$, yet they would remain negligible. To achieve an observable visibility drop (of order $0.01-1$), one would need a trajectory separation of $\Delta z\sim10-100\,\mathrm{cm}$. Such separations exceed current capabilities and would likely require two independent lattice lasers.

Note that the observation of the frequency shift~\eqref{frequency shift} itself would not be evidence of the quantum clock reading the superposition of proper times, since the frequency shift is the same as for the clock being in a statistical mixture of two heights. To reject the latter possibility, one can, apart from the total probability $P_g$ of measuring the atom in the ground state at the output, extract from the measured data also the difference $D_g\equiv P_g^{(0)}-P_g^{(1)}$ given by (compare with Eq.~\eqref{partial probability})
\begin{equation}
    D_g=\frac{1}{8}\cos\left(\Delta\phi+\frac{\delta_{\rm u}-\delta_{\rm d}}{2}\right)\sin\left(\frac{\delta_{\rm d}}{2}\right)\sin\left(\frac{\delta_{\rm u}}{2}\right),
\end{equation}
where $\delta_{\rm d}$ and $\delta_{\rm u}$ are given in Eq.~\eqref{phases}, and $\Delta\phi$ reads (see Appendix~\ref{supp:phases} for derivation)
\begin{equation}
    \Delta\phi=-\frac{mg\Delta z}{\hbar}\left(T+2T'+\tau\right).
\end{equation}
If the atom is in a superposition of heights, $D_g$ will oscillate with time~$T_{\rm B}$. On the other hand, for the atom following a statistical mixture of two heights, the probability difference would be equal to zero. Therefore, extracting from the interferometric output the fractional frequency shift and the oscillations of~$D_g$ would provide evidence that the atom was reading a superposition of proper times.

It is important to emphasize that the method proposed in this work does not rely on the assumption that the ground and excited states follow the same trajectories within the optical lattice. While we showed that, within the required order $1/c^2$ of calculating the phases acquired in the interferometer, the trajectories associated with different mass–energy states can be approximated as equal, the method allows accounting for mass–energy corrections and the resulting phases beyond this approximation. Furthermore, we emphasize that our results do not depend crucially on the simplified Bloch oscillation model we have adopted. Indeed, in Appendix~\ref{supp:Bloch oscillations}, we derive analogous results using the usual description of Bloch oscillations~\cite{Bloch1929, Zener1934}, in which the atom oscillates under a periodic potential of the optical lattice.

The assumption of zero momentum transfer from the clock pulse is well motivated in special cases, including the usual Lamb‐Dicke regime of optical-lattice traps~\cite{Dicke1953, Bergquist1987, Ido2003, Katori2003}, and is exact in the case of two‐photon E1-M1 clock transitions~\cite{Grynberg1983, Alden2014, Janson2024}. The Lamb-Dicke regime is reached when the atomic motion is limited to a region smaller than the transition wavelength. It was shown~\cite{Dicke1953} that in this case, most transitions occur at the internal transition frequency and do not alter the center-of-mass motion of the atom. However, such behavior is well understood only in the fully quantum picture (with quantized center-of-mass motion) --- in our simplified picture with the atom following a semiclassical trajectory, the no-recoil assumption is made as a simplifying yet justified approximation. By contrast, for E1-M1 transitions~\cite{Grynberg1983, Alden2014, Janson2024}, involving two counter-propagating photons with equal frequencies, the clock transition is recoilless as a rule. In this case, the no-recoil assumption is also well motivated in our simplified model of Bloch oscillations.

Abandoning the assumption of zero momentum transfer naturally leads to recoil shifts~\cite{Kolchenko1969, Hall1976}. In our model, this implies that the ground and excited states follow different trajectories, and the description becomes significantly more involved. The analysis of this case goes beyond the scope of the present work.

Note that the interpretation of the experiment as spectroscopy of a clock in a superposition of heights is possible only by including the laser phases and the explicit mechanism that establishes the quantum clock. These elements were absent in the idealized scenarios of~\cite{Sinha2011, Zych2011} and in the recent proposal using optical tweezers~\cite{Meltzer2024}. Consequently, the possibility of measuring the relativistic frequency shift in such interferometric setups has gone unnoticed until now.

\begin{acknowledgments}
We thank Timothy Le Pers, Navdeep Arya, Germain Tobar, Jun Ye, Kyungtae Kim, and Charles Baynham for helpful discussions and useful comments. The authors acknowledge the Knut and Alice Wallenberg Foundation through a Wallenberg Academy Fellowship No.\ 2021.0119. J.F.\ acknowledges funding from the U.S.\ Department of Energy, Office of Science, ASCR under Award Number DE-SC0023291 and Australian Research Council Centre of Excellence for Quantum Computation and Communication Technology (No.\ CE1701000012).
\end{acknowledgments}

\bibliographystyle{quantum}
\bibliography{apssamp}

\onecolumngrid

\appendix
%
%

\section{Probability calculation}\label{supp:probability calculation}
Here we derive the formula~\eqref{partial probability} for the probability $P_g^{(j)}$ of measuring the atom in the ground state in output $j\in\{0,1\}$. We denote by $\Delta\phi$ the phase difference between upper and lower ground state trajectories, and by $\delta_{\rm d}$ and $\delta_{\rm u}$ the additional phase acquired by the excited state on the lower and upper trajectory, respectively. The final state of the atom $\ket{\psi_\text{final}}$ can be written in a generic way
\begin{equation}
    \ket{\psi_\text{final}}=\frac{1}{8}\ket{g}\left(\ket{0}\left[1-\mathrm{e}^{\mathrm{i}\delta_{\rm d}}+\mathrm{e}^{\mathrm{i}\Delta\phi}\left(1-\mathrm{e}^{\mathrm{i}\delta_{\rm u}}\right)\right]-\mathrm{i}\ket{1}\left[1-\mathrm{e}^{\mathrm{i}\delta_{\rm d}}-\mathrm{e}^{\mathrm{i}\Delta\phi}\left(1-\mathrm{e}^{\mathrm{i}\delta_{\rm u}}\right)\right]\right)+\ldots,
\end{equation}
where $\ket{0}$ and $\ket{1}$ are the states corresponding to two relevant output trajectories, and dots at the end stand for the parts of the state irrelevant to our considerations (lost trajectories and excited state). The factor $1/8=(1/\sqrt{2})^6$ in front comes from six laser pulses applied to the atom (four Bragg pulses and two clock pulses). The minus signs in the bracket multiplying $\ket{0}$ result from the fact that excited state trajectories interacted with a laser two more times than the ground state trajectories (two clock pulses) --- each interaction gives rise to the factor $\mathrm{i}\mathrm{e}^{\mathrm{i}\phi_l}$ multiplying the corresponding trajectory (here $\phi_l$ is the so-called laser phase and we absorb it into phases $\Delta\phi$, $\delta_{\rm d}$, and $\delta_{\rm u}$). The factor $-\mathrm{i}$ multiplying $\ket{1}$ reflects the fact that trajectories leaving the interferometer in output 1 interacted with the odd number of Bragg pulses (the lower trajectory interacted once, the upper one three times), while the ones leaving in output 0 interacted with the even number (two) of Bragg pulses. Finally, the additional minus sign in the bracket multiplying $\ket{1}$ comes from the fact that, in the case of trajectories leaving in output 1, the upper trajectory interacted with two more Bragg pulses than the lower one.

The amplitude of finding the atom in the ground state in the output $j$ is given by
\begin{equation}
    \bra{g,j}\ket{\psi_\text{final}}\propto\frac{1}{8}\left[1-\mathrm{e}^{\mathrm{i}\delta_{\rm d}}+(-1)^j\mathrm{e}^{\mathrm{i}\Delta\phi}\left(1-\mathrm{e}^{\mathrm{i}\delta_{\rm u}}\right)\right],
\end{equation}
where we omitted the common phase factor $-\mathrm{i}$ present in the case of $j=1$. To calculate the probability $P_g^{(j)}$ we take the absolute value squared of the above amplitude and get
\begin{equation}
    P_g^{(j)}=\left|\bra{g,j}\ket{\psi_\text{final}}\right|^2=\frac{1}{16}\left[1-\cos\left(\frac{\delta_{\rm u}+\delta_{\rm d}}{2}\right)\cos\left(\frac{\delta_{\rm u}-\delta_{\rm d}}{2}\right)+(-1)^{j}2\cos\left(\Delta\phi+\frac{\delta_{\rm u}-\delta_{\rm d}}{2}\right)\sin\left(\frac{\delta_{\rm d}}{2}\right)\sin\left(\frac{\delta_{\rm u}}{2}\right)\right],
\end{equation}
which is exactly Eq.~\eqref{partial probability}.

\section{Fall of the excited-state trajectory}\label{supp:trajectory fall}
Let us motivate the use of the perturbative method~\cite{Storey1994} by analyzing the fall of the excited-state trajectory during Bloch oscillations. We assume that the frequencies of the optical lattice lasers are chosen such that the ground state follows a bouncing trajectory at a fixed height (i.e., each interaction with the laser occurs at the same height $z_{\rm B}$). Then, the excited-state trajectory necessarily falls with each bounce, due to its greater mass. Indeed, since we assume that the interaction with the laser occurs only when the atom reaches downward velocity $v_{\rm B}=\hbar k/m$, and the momentum transferred to the atom in each interaction is the same for both the ground and the excited state, the excited state will move after each interaction with the upward velocity $v_{\rm B}-\delta v_{\rm B}$, where the velocity difference $\delta v_{\rm B}$ is given by
\begin{equation}\label{velocity difference}
    \delta v_{\rm B}=2v_{\rm B}\frac{\delta m}{m}.
\end{equation}
Since the velocity of the excited state right before each interaction is larger (in terms of the absolute value) than right after it, each subsequent interaction will occur lower by
\begin{equation}\label{height difference}
    \delta z_{\rm B}=\frac{v_{\rm B}\delta v_{\rm B}}{g}
\end{equation}
than the previous one meaning that the excited-state trajectory will progressively fall.

Also, the excited trajectory's smaller starting velocity means it will reach the downward velocity $v_{\rm B}$ faster than the ground state. Two subsequent interactions on the excited-state trajectory are separated by a time $\tau_{\rm B}-\delta\tau_{\rm B}$, where
\begin{equation}\label{time difference}
    \delta \tau_{\rm B}=\frac{\delta v_{\rm B}}{g}.
\end{equation}
The spatial separation between the ground and the excited state trajectories is maximal when the excited state interacts with the laser. Then, it changes the direction of motion and meets the ground state trajectory at its interaction height $z_{\rm B}$. Therefore, the trajectories meet repeatedly (at the interaction points of the ground state trajectory), but at each subsequent meeting point the excited-state trajectory has velocity lower by $\delta v_{\rm B}$ than at the previous meeting point (while the ground state moves there at velocity $v_{\rm B}$).

The largest separation between the $n$th and $(n+1)$th interaction on the ground state trajectory occurs at the moment of $(n+1)$th interaction on the excited-state trajectory and is equal to $2n\delta z_{\rm B}$. For the experimental parameters from~\cite{Zhang2016} (strontium atoms, optical lattice operating at $532\;\text{nm}$, around 500 oscillations) the maximal separation between the ground and excited state trajectories would be around $10^{-13}\;\text{m}$. This should be compared with the thermal wavelength of the atoms given by
\begin{equation}
    \lambda_\text{th}=\sqrt{\frac{2\pi\hbar^2}{mk_\text{\tiny{B}}T}},
\end{equation}
where $T$ is the temperature. In the experiment~\cite{Zhang2016} the atoms were cooled to $\sim400\;\text{nK}$, corresponding to $\lambda_\text{th}\sim10^{-7}\;\text{m}$. Therefore, the thermal wavelength is much larger than the separation between the trajectories, and it is justified to use the perturbative approach in phase calculations.

\section{Phases}\label{supp:phases}
Let us calculate the phases $\Delta\phi$, $\delta_{\rm d}$, and $\delta_{\rm u}$ using the perturbative method developed in~\cite{Storey1994}. Each consists of two contributions: the propagation phase calculated by integrating the free-fall Lagrangian over the trajectory followed by the atom, and the laser phase resulting from atom-light interactions.

The free-fall Lagrangian of a relativistic particle with the internal two-level structure incorporating the mass-energy equivalence can be expanded to the order $1/c^2$ as follows:
\begin{equation}\label{correction 1}
\begin{split}
    L=-\hat{m}c^2-\hat{m}\left(gz-\frac{1}{2}v^2\right)-\frac{\hat{m}}{c^2}\left(\frac{1}{2}g^2z^2-\frac{1}{8}v^4+\frac{3}{2}gzv^2\right)+\mathcal{O}\left(\frac{1}{c^4}\right).
\end{split}
\end{equation}
Here $\hat{m}$ is the mass operator returning $m$ for the ground state, and $m+\delta m$ (where $\delta m=\hbar\omega_0/c^2$) for the excited state of the atom. At this point, we distinguish the $1/c^2$ correction to the excited-state Lagrangian coming from the mass-energy equivalence
\begin{equation}
    \delta L^\text{(1)}=-\delta m\left(gz-v^2/2\right),
\end{equation}
and the correction to the (ground- and excited-state) Lagrangian coming from the expansion of the relativistic-particle Lagrangian up to $1/c^2$ order
\begin{equation}
    \delta L^\text{(2)}=-\frac{m}{c^2}\left(\frac{1}{2}g^2z^2-\frac{1}{8}v^4+\frac{3}{2}gzv^2\right).
\end{equation}
Here we replaced $\hat{m}$ by $m$ because $\delta m$ is of the order $1/c^2$ and would contribute in higher order to the above expression. Note that for the scenario considered in this work, based on numbers from~\cite{Zhang2016}, $\delta L^{(2)}\ll\delta L^{(1)}$. This is because the internal energy of the atomic transition ($\sim1\;\text{eV}$) is much larger than the kinetic energy of the atom or the gravitational energy difference between the trajectories (both $\sim10^{-10}\;\text{eV}$).

Note also that the correction $\delta L^{(2)}$ and the corrections to the laser phases derived in~\cite{Werner2024} depend only on the trajectory, not on the internal state. Therefore, in the picture in which both the ground and the excited state follow the same trajectory, they will contribute only to the phase $\Delta\phi$, not to $\delta_{\rm d}$ and $\delta_{\rm u}$. We are not interested in $1/c^2$ corrections to $\Delta\phi$, hence we will omit them in the subsequent analysis and focus only on the corrections coming from $\delta L^{(1)}$.

We begin with the calculation of $\Delta\phi$. To calculate the propagation phase contribution we divide the trajectory into freely falling pieces between points of interaction with the lasers and calculate the propagation phase by integrating the freely falling Lagrangian
\begin{equation}
    L_g=-m\left(c^2+gz-v^2/2\right)
\end{equation}
along the trajectory. More precisely, for the free fall starting at time $t_\text{A}$ and ending at $t_{\rm B}$, the propagation phase is given by
\begin{equation}\label{propagation phase}
\begin{split}
    \phi_\text{p}(t_\text{A}\to t_{\rm B})=&\frac{1}{\hbar}\int_{t_\text{A}}^{t_{\rm B}}\mathrm{d}t L_g=-\frac{m}{\hbar}\int_{t_\text{A}}^{t_{\rm B}}\mathrm{d}t \left(c^2+gz-v^2/2\right)\\
    =&-\frac{m}{\hbar}\left[\left(c^2+gz_\text{A}-\frac{1}{2}v_\text{A}^2\right)(t_{\rm B}-t_\text{A})+v_\text{A}g(t_{\rm B}-t_\text{A})^2-\frac{1}{3}g^2(t_{\rm B}-t_\text{A})^3\right],
\end{split}
\end{equation}
where $z_\text{A}$ and $v_\text{A}$ are the atom's position and velocity, respectively, at time $t_\text{A}$. Denote by $t_j$ with $j\in\{1,2,3,4\}$ the times at which consecutive Bragg pulses are applied, and by $t_\text{i}$ and $t_\text{f}$ the times of the initial and final clock pulses, respectively. Notice that to match the notation from the main text, the times of application of particular pulses must satisfy
\begin{equation}
    t_2-t_1=t_4-t_3=T,\qquad t_\text{i}-t_2=t_3-t_\text{f}=T',\qquad t_\text{f}-t_\text{i}=T_{\rm B}.
\end{equation}
Let us further denote the quantities corresponding to the lower and upper trajectories by superscripts $\text{(d)}$ and $\text{(u)}$, respectively, and by $\Delta\phi_\text{p}(t_\text{A}\to t_{\rm B})=\phi_\text{p}^\text{(u)}(t_\text{A}\to t_{\rm B})-\phi_\text{p}^\text{(d)}(t_\text{A}\to t_{\rm B})$ the propagation phase difference between those two trajectories acquired between $t_\text{A}$ and $t_{\rm B}$. Let us consider one by one the consecutive stages of the interferometer and calculate the corresponding phase difference using~\eqref{propagation phase}.
\begin{itemize}
    \item $t_1\to t_2$:\newline
    At time $t_1$ both trajectories are at the same height, but the upper one starts with a velocity greater by $\Delta v$ than the lower one. Denote the initial velocity of the lower trajectory by $v_1$. Then, the relative propagation phase acquired between $t_1$ and $t_2$ reads
    \begin{equation}
        \Delta\phi_\text{p}(t_1\to t_2)=\frac{m}{\hbar}\Delta vT\left(v_1+\frac{1}{2}\Delta v-gT\right).
    \end{equation}
    \item $t_2\to t_\text{i}$:\newline
    Both trajectories start with the same velocity, but the upper one is higher by $\Delta v T$. The relative propagation phase is given by
    \begin{equation}
        \Delta\phi_\text{p}(t_2\to t_\text{i})=-\frac{m}{\hbar}g\Delta vTT'.
    \end{equation}
    \item $t_\text{i}\to t_\text{f}$:\newline
    We assume that both trajectories oscillate with the same frequency, and start simultaneously with zero initial velocity and maximal height. Therefore, the trajectories are constantly separated by $\Delta vT$ and have the same velocities all the time. The relative propagation phase equals
    \begin{equation}
        \Delta\phi_\text{p}(t_\text{i}\to t_\text{f})=-\frac{m}{\hbar}g\Delta vTT_{\rm B}.
    \end{equation}
    \item $t_\text{f}\to t_3$:\newline
    Similar to the previous two points, the velocity on both trajectories is the same, and they are separated by $\Delta vT$. The propagation phase difference reads
    \begin{equation}
        \Delta\phi_\text{p}(t_\text{f}\to t_\text{3})=-\frac{m}{\hbar}g\Delta vTT'.
    \end{equation}
    \item $t_3\to t_4$:\newline
    The trajectories start separated by $\Delta vT$, and the upper one has a velocity smaller by $\Delta v$. Denote the initial velocity of the lower trajectory by $v_3$. The relative propagation phase is then given by
    \begin{equation}
        \Delta\phi_\text{p}(t_3\to t_4)=\frac{m}{\hbar}\Delta vT\left(-v_3+\frac{1}{2}\Delta v\right).
    \end{equation}
\end{itemize}
To calculate the total propagation phase difference $\Delta\phi_\text{p}$ we sum up all the contributions, which gives
\begin{equation}\label{propagation phase difference 1}
    \Delta\phi_\text{p}=\frac{m}{\hbar}\Delta vT\left[v_1+\Delta v-v_3-g\left(T+2T'+T_{\rm B}\right)\right].
\end{equation}
Finally, let us note that, because of the requirement that at $t_\text{i}$ the velocity on both trajectories is equal to zero, the following relation holds
\begin{equation}\label{v1}
    v_1+\Delta v=g(T+T').
\end{equation}
On the other hand, $v_3$ is the velocity that the atom reaches after time $T'$ of a free fall after leaving the optical lattice. Let us notice that the atom does not leave the optical lattice with zero velocity (see Fig.~\ref{Fig::Bloch oscillations}), but rather with velocity $-g\tau$. Here $\tau\in\{-\tau_{\rm B}/2,\tau_{\rm B}/2\}$ is given by
\begin{equation}
    T_{\rm B}=N\tau_{\rm B}+\tau,
\end{equation}
where $N=\lfloor T_{\rm B}/\tau_{\rm B}+1/2\rfloor$ is the number of interactions between the optical lattice and the atom. Therefore, we have 
\begin{equation}\label{v3}
    v_3=-g(\tau+T')
\end{equation}
and we can rewrite Eq.~\eqref{propagation phase difference 1} (introducing $\Delta z=\Delta vT$) as follows:
\begin{equation}\label{propagation phase difference 2}
    \Delta\phi_\text{p}=\frac{m}{\hbar}\Delta zg(\tau-T_{\rm B})=-\frac{m}{\hbar}g\Delta zN\tau_{\rm B}.
\end{equation}

Now, let us focus on the laser phase contribution to the phase $\Delta\phi$. As described in~\cite{Storey1994}, when the atom absorbs (emits) a photon with frequency $\omega$ and wave vector $k$, it acquires a laser phase
\begin{equation}\label{laser phase}
    \phi_\text{l}(t,z)=\mp\omega t\pm kz,
\end{equation}
where the upper (lower) sign corresponds to photon absorption (emission), and $t$ and $z$ are the time and position of the interaction, respectively. In the case of Bragg pulses, as well as in the optical lattice, the photon absorption is immediately followed by the emission of a photon with the same frequency and opposite wave vector. Therefore, the laser phase acquired at each such interaction point is equal to $\phi_\text{l}(t,z)=\phi_\text{l}(z)=\pm2kz$ with the plus (minus) sign corresponding to the atom receiving upward (downward) momentum.

Denote by $z_j$ with $j\in\{1,2,3,4\}$ the interaction heights with four consecutive Bragg pulses ($z_1$ and $z_3$ correspond to the upper trajectory, while $z_2$ and $z_4$ to the lower one), and by $z_{\rm d}$ and $z_{\rm u}$ the heights of interactions of the lower and upper trajectories, respectively, within the optical lattice. Note also that the wave vector $k_\text{Bragg}$ of the pulses is related to the velocity change $\Delta v$ by
\begin{equation}\label{Bragg}
    2\hbar k_\text{Bragg}=m\Delta v,\qquad\Longrightarrow\qquad k_\text{Bragg}=\frac{m\Delta v}{2\hbar},
\end{equation}
and the wave vector $k_\text{Bloch}$ of the optical lattice laser is related to the Bloch period $\tau_{\rm B}$ by
\begin{equation}\label{Bloch}
    2\hbar k_\text{Bloch}=mg\tau_{\rm B},\qquad\Longrightarrow\qquad k_\text{Bloch}=\frac{mg\tau_{\rm B}}{2\hbar}.
\end{equation}
The total laser phase difference (between the upper and lower trajectory) is given by
\begin{equation}
    \Delta\phi_\text{l}=2k_\text{Bragg}(z_1-z_2-z_3+z_4)+2Nk_\text{Bloch}(z_{\rm u}-z_{\rm d})
\end{equation}
Let us notice the following relations between particular heights:
\begin{equation}
    z_2-z_1=v_1T-\frac{1}{2}gT^2,\qquad z_4-z_3=(v_3-\Delta v)T-\frac{1}{2}gT^2,\qquad z_{\rm u}-z_{\rm d}=\Delta z,
\end{equation}
where $v_1$ and $v_3$ are given by \eqref{v1} and \eqref{v3}. Hence, we can write the total laser phase difference as
\begin{equation}\label{total laser phase}
    \Delta\phi_\text{l}=\frac{m}{\hbar}\left[-g\Delta z(T+2T'+\tau)+g\Delta zN\tau_{\rm B}\right].
\end{equation}
The total phase difference $\Delta\phi$ between the upper and the lower (ground state) trajectories is calculated by summing up the total propagation phase difference and the total laser phase difference. This gives
\begin{equation}
    \Delta\phi=\Delta\phi_\text{p}+\Delta\phi_\text{l}=-\frac{mg\Delta z}{\hbar}(T+2T'+\tau).
\end{equation}

Now, let us calculate the additional phases $\delta_{\rm d}$ and $\delta_{\rm u}$ acquired by the excited state on the lower and upper trajectory, respectively. We need to calculate the additional laser phase coming from the clock pulses and the correction to the propagation phase due to the mass-energy equivalence. Starting with the laser phase, we assume that the clock pulses change only the atom's internal state, but do not affect its motion. This can be achieved by driving two-photon transitions with both photons having the same frequency $\omega/2$ and opposite momenta. The corresponding laser phase is equal to $\mp\omega t$ with the minus (plus) sign corresponding to the two-photon absorption (emission). Note that this phase is independent of the position $z$ of the interaction (thus, it is the same for the lower and upper trajectory). Since the excited state trajectories absorb the photons at $t_\text{i}$ and emit at $t_\text{f}$, they acquire the laser phase $\omega(t_\text{f}-t_\text{i})=\omega T_{\rm B}$.

To calculate the propagation phase correction, we employ the perturbative method developed in~\cite{Storey1994} and integrate the mass-energy correction to the free-fall Lagrangian
\begin{equation}
    \delta L=-\delta m\left(c^2+gz-v^2/2\right)
\end{equation}
over the ground state trajectory on the corresponding height. For the free-fall trajectory starting at $t_\text{a}$ and ending at $t_{\rm B}$ the correction $\delta_\text{p}(t_\text{a}\to t_{\rm B})$ to the propagation phase reads (compare with~\eqref{propagation phase})
\begin{equation}
    \delta_\text{p}(t_\text{a}\to t_{\rm B})=-\frac{\delta m}{\hbar}\left[\left(c^2+gz_\text{a}-\frac{1}{2}v_\text{a}^2\right)(t_{\rm B}-t_\text{a})+v_\text{a}g(t_{\rm B}-t_\text{a})^2-\frac{1}{3}g^2(t_{\rm B}-t_\text{a})^3\right].
\end{equation}
For a full single oscillation in the optical lattice starting at height $z_{\rm B}$ with velocity $v_{\rm B}$, the propagation phase correction $\delta_\text{p,1}$ is given by
\begin{equation}\label{single bounce propagation phase}
    \delta_\text{p,1}=\frac{\delta m}{\hbar}\left[-c^2-g\left(z_{\rm B}+\frac{v_{\rm B}^2}{3}\right)+\frac{v_{\rm B}^2}{6}\right]\tau_{\rm B}\equiv-\omega_0\tau_{\rm B}\left[1+\frac{g\langle z\rangle}{c^2}-\frac{\langle v^2\rangle}{2c^2}\right],
\end{equation}
where $\tau_{\rm B}$ is the oscillation period, we have used the fact that $\delta m=\hbar\omega_0/c^2$, and introduced $\langle z\rangle$ and $\langle v^2\rangle$ for the mean height and velocity squared. We assume that the total number $N$ of Bloch oscillations is large, and neglect the correction coming from the fact that the total time $T_{\rm B}$ of Bloch oscillations is not necessarily equal to the integer multiple of Bloch periods (instead, $T_{\rm B}=N\tau_{\rm B}+\tau$ with $\tau\in(-\tau_{\rm B}/2,\tau_{\rm B}/2)$; however, since $\tau\ll N\tau_{\rm B}$, we will assume $\tau\approx0$). Then, the total propagation phase correction is given by
\begin{equation}
    \delta_\text{p,tot}\approx N\delta_\text{p,1}=-\omega_0N\tau_{\rm B}\left[1+\frac{g\langle z\rangle}{c^2}-\frac{\langle v^2\rangle}{2c^2}\right]\approx-\omega_0T_{\rm B}\left[1+\frac{g\langle z\rangle}{c^2}-\frac{\langle v^2\rangle}{2c^2}\right].
\end{equation}

Without loss of generality, we can assume that the mean height corresponding to the lower trajectory $\langle z\rangle_{\rm d}=0$, while this of the upper one equals $\langle z\rangle_{\rm u}=\Delta z$. Combining the propagation phase with the laser phase corrections, we get
\begin{equation}\label{phases result}
    \begin{split}
        &\delta_{\rm d}=\left[\omega-\omega_0\left(1-\frac{\langle v^2\rangle}{2c^2}\right)\right]T_{\rm B},\\
        &\delta_{\rm u}=\left[\omega-\omega_0\left(1+\frac{g\Delta z}{c^2}-\frac{\langle v^2\rangle}{2c^2}\right)\right]T_{\rm B}.
    \end{split}
\end{equation}
%

\section{Non-perturbative calculations}\label{supp:non-perturbative calculations}

As a consistency check let us compute the phases accumulated during the Bloch oscillations non-perturbatively, namely for the trajectories described in Appendix~\ref{supp:trajectory fall}, and make sure that they agree with the perturbative calculation from Appendix~\ref{supp:phases}. In the following, we restrict all the calculations to the terms up to order $1/c^2$ neglecting the higher-order terms. This is consistent with the perturbative approach, where we considered only $1/c^2$ corrections.

Since the trajectories of the ground and excited states no longer coincide (the excited trajectory falls), the phase difference between them will gain contribution from three different sources: the propagation phase, the laser phase, and the separation phase. The first two are defined in the same way as in the previous (perturbative) considerations, and the third one is given by
\begin{equation}\label{separation phase definition}
    \delta_\text{s}(t)\equiv\frac{\bar{p}(t)\Delta z(t)}{\hbar},
\end{equation}
where $\bar{p}(t)$ is the mean momentum of two trajectories at time $t$ (the oscillation starts at $t=0$), and $\Delta z(t)$ is the separation between them (note that this has nothing to do with the separation $\Delta z$ between the upper and lower trajectory).

Since the phase acquired due to the interaction with the clock pulses is position-independent, it will not change in the non-perturbative analysis. Therefore, we consider only the phases acquired during the Bloch oscillations within the optical lattice. The trajectory followed by both trajectories has been described above. We analyze only the time between the two first interactions on the ground state trajectory (at some arbitrary height $z_{\rm B}$) and show that the phase difference between the ground and the excited state trajectories agrees with the perturbative calculations. It is straightforward that it will agree with the perturbative calculations also at later times.

The propagation phases acquired on both trajectories in the first part of the oscillation (before the second interaction of the excited state) can be calculated using the general formula for the propagation phase in free fall:
\begin{equation}
    \phi_\text{p}^\text{(g/e)}(t)=\frac{m^\text{(g/e)}}{\hbar}\left[\left(-c^2+\frac{1}{2}v_0^2-gz_0\right)t-v_0gt^2+\frac{1}{3}g^2t^3\right],
\end{equation}
(this is just Eq.~\eqref{propagation phase} with $t_\text{A}\to0$ and $t_{\rm B}\to t$) where $m^\text{(g)}=m$ and $m^\text{(e)}=m+\delta m$ are the masses of the atom in the ground and excited state, respectively, $v_0$ is the initial velocity, and $z_0$ is the initial height. For the ground state, we replace $v_0$ by $v_{\rm B}$, while for the excited state by $v_{\rm B}-\delta v_{\rm B}$. The initial height for both states is the same and equal to $z_{\rm B}$.

The difference in propagation phases at time $t$ is then given by
\begin{equation}
    \delta_\text{p}(t)=\phi_\text{p}^\text{(e)}(t)-\phi_\text{p}^\text{(g)}(t)=\frac{\delta m}{\hbar}\left[\left(-c^2+\frac{1}{2}v_{\rm B}^2-gz_{\rm B}\right)t-v_{\rm B}gt^2+\frac{1}{3}g^2t^3\right]-\frac{m}{\hbar}(v_{\rm B}-gt)\delta v_{\rm B}t
\end{equation}
(recall that we neglect the terms higher-order than $1/c^2$). Let us notice that, since $\delta v_{\rm B} t$ is just the separation $\Delta z(t)$ between the trajectories, the second term in the above formula is equal in value but with the opposite sign to the separation phase
\begin{equation}
    \delta_\text{s}(t)=\frac{\bar{p}(t)\delta v_{\rm B} t}{\hbar}=\frac{m}{\hbar}(v_{\rm B}-gt)\delta v_{\rm B}t,
\end{equation}
where $\bar{p}(t)=m(v_{\rm B}-gt)+\mathcal{O}(1/c^2)$. Therefore, the total phase difference for $t<\tau_{\rm B}-\delta v_{\rm B}/g$ is given by
\begin{equation}
    \delta_\text{p}(t)+\delta_\text{s}(t)=\frac{\delta m}{\hbar}\left[\left(-c^2+\frac{1}{2}v_{\rm B}^2-gz_{\rm B}\right)t-v_{\rm B}gt^2+\frac{1}{3}g^2t^3\right],
\end{equation}
in agreement with the perturbative results (indeed, this is just the propagation phase correction calculated in the perturbative approach).

At the time $\tau_{\rm B}-\delta v_{\rm B}/g$ the interaction of the atom on the excited trajectory with the laser occurs, and it acquires a laser phase
\begin{equation}
    \phi_\text{l}^\text{(e)}=\frac{m}{\hbar}2v_{\rm B}(z_{\rm B}-\delta z_{\rm B})=\phi_\text{l}^\text{(g)}-\frac{m}{\hbar}\delta v_{\rm B}\tau_{\rm B}+\mathcal{O}(\delta m^2),
\end{equation}
where $\phi_\text{l}^\text{(g)}$ is the laser phase acquired by the ground state trajectory at time $\tau_{\rm B}$. Let us notice that right before the second interaction of the excited state the separation phase is equal to
\begin{equation}
    \delta_\text{s}(\tau_{\rm B}-\Delta v/g-\varepsilon)=-\frac{m}{\hbar}v_{\rm B}\delta v_{\rm B}t,
\end{equation}
while right after the interaction it is $\sim\mathcal{O}(1/c^4)$, because the mean momentum is then $\sim\mathcal{O}(1/c^2)$. This situation (mean momentum $\sim\mathcal{O}(1/c^2)$ and thus negligible value of the separation phase) persists until the second interaction on the ground state trajectory, at which time both trajectories are at the same height, but the excited one has the velocity $v_{\rm B}-2\delta v_{\rm B}$. At this point, we can repeat the entire analysis changing everywhere $\delta v_{\rm B}\to2\delta v_{\rm B}$. Since the propagation phase difference acquired between the second interaction on the excited and on the ground trajectories is also $\sim\mathcal{O}(\delta m^2)$, the total phase difference after the first oscillation is equal to
\begin{equation}
    \delta=\delta_\text{p}+\delta_\text{s}+\delta_\text{l}=\frac{\delta m}{\hbar}\left[\left(-c^2+\frac{1}{2}v_{\rm B}^2-gz_{\rm B}\right)\tau_{\rm B}-v_{\rm B}g\tau_{\rm B}^2+\frac{1}{3}g^2\tau_{\rm B}^3\right]=\frac{\delta m}{\hbar}\left[-c^2-g\left(z_{\rm B}+\frac{v_{\rm B}^2}{3}\right)+\frac{v_{\rm B}^2}{6}\right]\tau_{\rm B},
\end{equation}
which is equal to the propagation phase correction calculated for a single oscillation in the perturbative approach (compare with Eq.~\eqref{single bounce propagation phase}).

\section{Bloch oscillations --- the usual description}\label{supp:Bloch oscillations}
In our analysis, we have employed a simplified model of Bloch oscillations based on~\cite{Cadoret2009, Andia2013}, where the oscillating atom follows a bouncing trajectory in a gravitational field (see Fig.~\ref{Fig::Bloch oscillations}). However, this is not the most common description of Bloch oscillations --- usually one assumes that the dynamics of the atom is set by a periodic potential (optical lattice) and a linear correction (gravitational field), which leads to the oscillatory motion~\cite{Bloch1929, Zener1934}. The full Hamiltonian of the system involving the leading relativistic corrections reads
\begin{equation}\label{Hamiltonian}
    \hat{H}=mc^2+\frac{\hat{p}^2}{2m}+mg\hat{z}+V_0\cos(2k_\text{Bloch}\hat{z})+\hbar\omega_0\ket{e}\bra{e}\left(1+\frac{g\hat{z}}{c^2}-\frac{\hat{p}^2}{2m^2c^2}\right)\equiv\hat{H}^{(0)}+\hat{H}^{(1)},
\end{equation}
where $V_0$ is the depth of the optical lattice potential, $k_\text{Bloch}$ is the wave vector of the lattice photons, $\ket{e}\bra{e}$ is a projector on the excited state of the atom, and
\begin{equation}
    \begin{split}
        &\hat{H}^{(0)}=\hat{H}^{(0)}_\text{cm}+\hat{H}^{(0)}_\text{int},\\
        &\hat{H}^{(0)}_\text{cm}=\frac{\hat{p}^2}{2m}+mg\hat{z}+V_0\cos(2k_\text{Bloch}\hat{z}),\\
        &\hat{H}^{(0)}_\text{int}=mc^2+\hbar\omega_0\ket{e}\bra{e},\\
        &\hat{H}^{(1)}=\frac{\hbar\omega_0}{c^2}\ket{e}\bra{e}\left(g\hat{z}-\frac{\hat{p}^2}{2m^2}\right).
    \end{split}
\end{equation}

Writing Eq.~\eqref{Hamiltonian} we have assumed that the lattice operates at the ``magic'' wavelength~\cite{Katori2003, Ye2008}, which means that the harmonic potential experienced by the ground and excited state atoms is the same. Let us notice that the frequency of the light as perceived by the atom depends on its position and velocity (due to the relativistic Doppler shifts). Therefore, in principle, it can only be at the ``magic'' wavelength for an atom at some specific height and velocity. The atom at some other height, or moving with a different velocity, will perceive the light as shifted from this wavelength, leading to a dependence of the optical lattice potential on the center-of-mass degrees of freedom. Note that this is exactly analogous to the situation in optical lattice clocks, where the atoms are not perfectly static and are placed at different heights. However, it was shown in~\cite{Katori2003} that around the ``magic'' wavelength the potential changes very slowly with the frequency of the lattice laser, and the corrections due to Doppler shifts can be made negligible. Consequently, we shall omit them from further consideration.

To calculate the phase shifts in the Hamiltonian formalism we use the time-independent perturbation theory. We treat the Hamiltonian $\hat{H}^{(1)}$ in Eq.~\eqref{Hamiltonian} as a small correction to $\hat{H}^{(0)}$. The eigenstates of $\hat{H}^{(0)}$ belong to the Hilbert space $\mathcal{H}_\text{cm}\otimes\mathcal{H}_\text{int}$ with $\mathcal{H}_\text{cm}$ spanned by the Wannier-Stark states~\cite{Wannier1960}, and $\mathcal{H}_\text{int}$ by $\{\ket{g},\ket{e}\}$. We assume that the center-of-mass state (the part corresponding to $\mathcal{H}_\text{cm}$) is the same for both internal states of the atom.  In the zeroth order, the phase shift is equal on both trajectories (lower and upper) and reads
\begin{equation}
    \delta_{\rm d}^{(0)}=\delta_{\rm u}^{(0)}=(\omega-\omega_0)T_{\rm B}.
\end{equation}
The first-order (in $1/c^2$) correction to the phases results from the correction to the eigenenergies and is given by
\begin{equation}
        \delta_i^{(1)}=-\frac{1}{\hbar}\left(\left\langle \hat{H}^{(1)}\right\rangle_{e,i}-\left\langle \hat{H}^{(1)}\right\rangle_{g,i}\right)T_{\rm B},
\end{equation}
where the expectation values are evaluated on unperturbed states. In the ground state, the expectation value of $\hat{H}^{(1)}$ is zero irrespectively of the path, and in the excited state we have
\begin{equation}
    \left\langle \hat{H}^{(1)}\right\rangle_{e,i}=\frac{\hbar\omega_0}{c^2}\left(g\langle \hat{z}\rangle_i-\frac{\langle \hat{p}^2\rangle_i}{2m^2}\right).
\end{equation}
This leads to the total phase shift of the form:
\begin{equation}\label{phase shift Hamiltonian approach}
    \delta_i=\delta_i^{(0)}+\delta_i^{(1)}=\left[\omega-\omega_0\left(1+\frac{g\langle \hat{z}\rangle_i}{c^2}-\frac{\langle \hat{p}^2\rangle_i}{2m^2c^2}\right)\right]T_{\rm B},
\end{equation}
which, upon identifications $\langle \hat{z}\rangle_i$ and $\langle \hat{p}^2\rangle_i=m^2\langle v^2\rangle_i$, takes the same form as derived in Appendix~\ref{supp:phases}. Note the different meaning of the brackets in~\eqref{phase shift Hamiltonian approach} and those present e.g. in~\eqref{phases result} --- in the first case they indicate averaging over the quantum state, while in the latter they mean taking the average over the classical trajectory.

The same results can be derived in the Lagrangian formalism using the method developed in~\cite{Storey1994}. We assume that the atom follows a superposition of classical trajectories set by the (ground-state) Lagrangian
\begin{equation}\label{Bloch Lagrangian}
    L_g=-mc^2+\frac{1}{2}mv^2-mgz-V_0\cos(2k_\text{Bloch}z),
\end{equation}
corresponding to the Hamiltonian~\eqref{Hamiltonian}. Let us consider (following the reasoning from Appendix~\ref{supp:phases}) the dominant correction to the Lagrangian of the excited state atoms, i.e.,
\begin{equation}
    \delta L=-\delta m\left(c^2+gz-v^2/2\right).
\end{equation}
To calculate the phases $\delta_{\rm d}$ and $\delta_{\rm u}$ ($\Delta\phi$ does not depend on the model of Bloch oscillations that we use) we integrate the Lagrangian correction $\delta L$ over the trajectory $z(t)$ followed by the atom (assuming that the trajectory is the same for both internal states). This gives (recall that $\delta m=\hbar\omega_0/c^2$)
\begin{equation}
    \frac{1}{\hbar}\int_0^{T_{\rm B}}\mathrm{d}t\delta L=-\omega_0T_{\rm B}\left[1+\frac{g}{c^2}\left(\frac{1}{T_{\rm B}}\int_0^{T_{\rm B}}\mathrm{d}tz(t)\right)-\frac{1}{2c^2}\left(\frac{1}{T_{\rm B}}\int_0^{T_{\rm B}}\mathrm{d}tv^2(t)\right)\right]\equiv-\omega_0T_{\rm B}\left[1+\frac{g\langle z \rangle}{c^2}-\frac{\langle v^2 \rangle}{2c^2}\right],
\end{equation}
where we have introduced
\begin{equation}
    \langle z \rangle\equiv\frac{1}{T_{\rm B}}\int_0^{T_{\rm B}}\mathrm{d}tz(t),\qquad \langle v^2 \rangle\equiv\frac{1}{T_{\rm B}}\int_0^{T_{\rm B}}\mathrm{d}tv^2(t).
\end{equation}
Assuming that the mean height of the lower trajectory is set to zero and that of the upper one to $\Delta z$, that the mean squared velocity is the same on both heights and adding a laser phase $\omega T_{\rm B}$ coming from the interaction with the clock pulse, we get
\begin{equation}
    \begin{split}
        &\delta_{\rm d}=\left[\omega-\omega_0\left(1-\frac{\langle v^2\rangle}{2c^2}\right)\right]T_{\rm B},\\
        &\delta_{\rm u}=\left[\omega-\omega_0\left(1+\frac{g\Delta z}{c^2}-\frac{\langle v^2\rangle}{2c^2}\right)\right]T_{\rm B},
    \end{split}
\end{equation}
which is the same as derived in Appendix~\ref{supp:phases}. It is worth noting that the method developed in~\cite{Storey1994} was derived for quadratic Lagrangians, while Lagrangian~\eqref{Bloch Lagrangian} is nonquadratic. However, the compatibility with the result obtained in the Hamiltonian formalism suggests that the method from~\cite{Storey1994} is still applicable in our case.

\end{document}